%% file: main.tex
\documentclass[runningheads]{llncs}
\usepackage[T1]{fontenc}
\usepackage{graphicx}
\usepackage{caption,float}
\usepackage{pgf} 
\usepackage{pgfplots} 
\usepackage{subcaption}
\usepackage{xspace}
\usepackage{booktabs}
\usepackage{enumitem}
\usepackage{cite}
\usepackage{xspace}
\usepackage{booktabs}
\usepackage{hyperref}
\usepackage{verbatim}
\usepgfplotslibrary{groupplots}
\pgfplotsset{compat=1.18} 

\usepackage[colorinlistoftodos]{todonotes}

\begin{document}
\title{CNN architecture extraction on edge GPU}
%
%

%
%

\author{P\'eter Horv\'ath\inst{1} \and
Lukasz Chmielewski\inst{2} \and
Leo Weissbart\inst{1} \and
Lejla Batina\inst{1} \and
Yuval Yarom\inst{3}
}
\authorrunning{P. Horv\'ath et al.}
%
\institute{Radboud University, Nijmegen, Netherlands \and
Masaryk University, Brno, Czech Republic \and
Ruhr University, Bochum, Germany
}

\maketitle              
\begin{abstract}
Neural networks have become popular due to their versatility and state-of-the-art results in many applications, such as image classification, natural language processing, speech recognition, forecasting, etc. These applications are also used in resource-constrained environments such as embedded devices.
In this work, the susceptibility of neural network implementations to reverse engineering is explored on the NVIDIA Jetson Nano microcomputer via side-channel analysis.
To this end, an architecture extraction attack is presented.
In the attack, 15 popular convolutional neural network architectures (EfficientNets, MobileNets, NasNet, etc.) are implemented on the GPU of Jetson Nano and the electromagnetic radiation of the GPU is analyzed during the inference operation of the neural networks. 
The results of the analysis show that neural network architectures are easily distinguishable using deep learning-based side-channel analysis.

\keywords{Deep Learning \and Side Channel Attack \and NVIDIA GPU.}
\end{abstract}
\section{Introduction}
\label{introduction}
The field of machine learning has seen an enormous amount of interest and use in recent years. One specific area of machine learning, namely deep learning, has proven to be versatile and provides state-of-the-art performance for many real-world applications. Deep learning refers to multi-layer Artificial Neural Networks (ANNs) that learn to solve a task by extracting the important features from data and generalizing well to that task. These tasks, such as games, object detection, image classification, or natural language processing, can be vastly different.

AlphaGo \cite{silver2016mastering} is one of the deep learning-based breakthroughs where a neural network learned to play Go and was able to beat one of the best human Go players at the time. Similarly, AlphaZero \cite{silver2017mastering} was developed to play chess and outperformed human players. Lastly, AlphaStar \cite{vinyals2019grandmaster} achieved superior levels, compared to human players, in the StarCraft 2 real-time strategy computer game, beating multiple of the best players in the world.

Additionally, image classification is a fundamental problem of computer vision where deep learning models have achieved state-of-the-art results and continue to provide improvements~\cite{krizhevsky2012imagenet, simonyan2014very, lin2013network, chollet2017xception, he2016deep}.
A more general problem in computer vision concerns the field of object detection, which has also seen enormous improvements in accuracy due to neural networks~\cite{liu2020deep}. 

Similarly, deep learning provided multiple breakthroughs in Natural Language Processing (NLP) in recent years~\cite{otter2020survey}. NLP is a broad field that aims to solve practical issues concerning human languages, such as information retrieval, summarisation, or machine translation. Google Translate~\cite{googletranslate} and ChatGPT~\cite{openai2023gpt4} are popular NLP applications based on the Transformer~\cite{vaswani2017attention} neural network architecture.

Neural networks are changing many areas of our lives and are becoming indispensable in our everyday lives.
However, the design and training of neural networks can be expensive in many ways, as follows:

\begin{enumerate}
    \item Collecting the training dataset can be time-consuming and expensive;
    \item Designing and training neural networks requires people with expertise;
    \item The time it takes to train and validate a model can range from hours to weeks;
    \item The cost of training and tuning can be high due to requiring specialized, high-performance hardware, e.g., graphics processing units (GPUs). 
\end{enumerate}

Additionally, sometimes sensitive data is used to train a neural network, and this data can also be vulnerable to reverse engineering.
Therefore, keeping the architecture and parameters of the trained models secret becomes an important issue.

Beyond their great successes, neural networks also face a wide variety of adversarial attacks. These attacks can have different goals, such as causing misclassification, input recovery, or reverse engineering architecture.

One kind of technique that attackers can employ is Side-Channel Analysis (SCA). SCA exploits the physical leakages of electronic devices to extract secret information. Despite existing countermeasures against SCA-based attacks, it is not always possible to utilize these countermeasures, especially in resource-constrained environments, because countermeasures usually come at a price of speed and cost.   

Therefore, this work will focus on the following research question:
\textbf{Are neural network implementations of large-scale convolutional neural networks on the GPU of NVIDIA Jetson Nano vulnerable to reverse engineering using deep learning-based side-channel analysis?}

The GPU platform is targeted because neural network implementations, especially convolutional neural networks, in practice, often run on GPUs because the core operations of neural networks are matrix operations (e.g., multiplications). These operations are highly parallelizable, which makes GPUs more suitable than CPUs for neural network-based applications.

\subsection{Comparison with related work}

In this work, we analyze electromagnetic and timing side channel information similarly to the CSI-NN paper, which analyzed NNs running on microcontrollers \cite{batina2019csi}. 
However, we focus on well-known and widely used, large-scale convolutional neural network architectures in computer vision with a different and more popular platform for neural networks as the target, i.e. the GPU. 
In practice, since GPUs provide efficiency through parallelism, they are commonly used to run  
neural networks, especially large-scale ones. 
This GPU parallelism also poses a big challenge in analyzing side-channel signals, as the number of concurrently executing threads is much larger than that of microcontroller applications.
Moreover, the attacks presented in this paper do not require the decapsulation of the target chip, contrary to~\cite{batina2019csi}. 

The work of Chmielewski and Weissbart~\cite{chmielewski2021reverse} targets the same platform as we do in this work, with the goals and methods similar to those of~\cite{batina2019csi}. Basically, the number of neurons, number of layers, and activation function types are recovered based on using electromagnetic side-channel and timing information. However, our work goes further in extending the approach to large-scale architectures and showing that recovering neural net architectures used in real-world application is viable. 

In addition, some other works target desktop GPUs to extract hyperparameters ~\cite{DBLP:conf/uss/MaiaXLGZ22, clairvoyance} but not an embedded system like the Jetson Nano that might be deployed in an environment where an adversary is more likely to have physical access.

A side-channel-based attack on neural networks is also presented in~\cite{9000972}, where the architecture extraction of neural networks, implemented on the CPU of Raspberry Pi, is demonstrated using power-side channel analysis and machine learning. The extracted architectures are similar to those in this work, but the classification method, the target platform, and the used side channel are different as they classify power traces with a Support Vector Machine (SVM) classifier. 

\subsection{Contributions and outline}

The target device in this work is the NVIDIA Jetson Nano, which is a microcomputer tailored to run AI applications in a resource-constrained environment. As already stated above, we demonstrate an architecture extraction attack by distinguishing among a number of well-know neural net architectures on this platform. 

To summarize, the main contributions of this work are:
\begin{enumerate}
    \item We demonstrate how complex convolutional neural network architectures can be extracted by visually inspecting electromagnetic side-channel measurements. To that end, 15 well-known neural network architectures from computer vision are classified based on the electromagnetic radiation of the device's GPU.
    \item We also show how the process of distinguishing the architectures can be automated using a deep learning classifier. 
\end{enumerate}

This papers is organized as follows.
Prior to discussing the experiments and results, Section \ref{background} gives an introduction to related topics. First, the investigated neural network architectures are discussed in detail. Next, the relevant concepts from side-channel analysis are introduced. Section \ref{archext} discusses the experimental setup as well as the results of reverse engineering. Section \ref{discussion} provides a discussion about the results and possible countermeasures. Section \ref{conclusions} concludes the paper.

\section{Background}
\label{background}

\subsection{CNN architectures}
This section introduces the convolutional neural network architectures that are analyzed in this work. Most of these architectures are suitable for resource-limited devices, such as the Jetson Nano, but there are other well-known architectures besides the ones analyzed in this work, such as ResNets \cite{he2016deep}, ShuffleNets \cite{zhang2018shufflenet} and Xception \cite{chollet2017xception}.

\subsubsection*{MobileNet}
MobileNets~\cite{howard2017mobilenets} are convolutional neural networks suitable for real-time applications in constrained environments. 
The architecture relies on depthwise separable convolutional blocks to speed up computations. 
These blocks consist of a depthwise convolutional layer and a pointwise convolutional layer. 
First, the depthwise convolutional layer applies $3 \times 3$ kernels on only one input channel of the input. 
Second, the produced feature map of the depthwise convolutional layer is the input to the pointwise convolutional layer with $1x1$ kernels, which are applied to all input channels. 
In standard convolutions, the $3 \times 3$ kernels would be applied to all input channels. 
Empirically, it has been shown that depthwise separable convolutions provide less latency with a negligible decrease in accuracy compared to standard convolutional layers. This is very important for embedded systems as the resources such as area and power consumption are typically limited.

\subsubsection*{MobileNetV2}
MobileNetV2~\cite{sandler2018mobilenetv2} is an optimized version of MobileNets. 
In this architecture, the depthwise separable convolutional blocks are expanded with linear bottleneck layers and residual connections~\cite{he2016deep}~\cite{xie2017aggregated} to form \textit{inverted residual blocks}.

\subsubsection*{EfficientNets}
EfficientNet~\cite{tan2019efficientnet} proposes a compound scaling method that uniformly scales model depth, width, and resolution with scaling coefficients. 
This compound scaling method is based on the intuition that all dimensions of a network have to be balanced to achieve better accuracy and efficiency. 
The baseline network, EfficientNetB0, is similar to that of MobileNetV2 as it is based on the same inverted residual blocks with bottleneck layers. 
In addition, squeeze-and-excitation~\cite{hu2018squeeze} is added to the blocks. 
The upscaled versions of the baseline architecture, such as EfficientNetB1, -B2, -B3, -B4, -B5, and -B6, are scaled up using the compound scaling method mentioned earlier.

\subsubsection*{DenseNets}
DenseNets~\cite{huang2017densely} do not use depthwise separable convolutions, they are based on the idea of feature reusing. 
In terms of the architecture, this means that feature maps produced by a layer are inputs to all subsequent layers. 
The architecture of DenseNets employs dense blocks and transition layers. 
Dense blocks use the principle of feature reusing, while transition layers are responsible for downsampling.

\subsubsection*{NasNetMobile}
The NasNet~\cite{zoph2018learning} architecture is quite distinct when compared to the previous architectures. 
The main building blocks of the NasNet architecture are the \textit{normal} and \textit{reduction} cells. 
These cells have multiple branches that apply different operations on the inputs in parallel,
and the results of the branches are concatenated to form the output of the cell. The operations in the branches consist of standard convolution, separable convolutions,
pooling or identity.

\subsubsection*{MobileNetV3}
MobileNetV3~\cite{howard2019searching} is the further optimized version of MobileNetV2 with various new additions. 
Similarly to MobileNetV2, MobileNetV3's main building blocks are the inverted residual blocks with bottleneck layers, 
but with the addition of squeeze-and-excitation~\cite{hu2018squeeze} in some blocks in the new architecture. 
Additionally, the ReLU nonlinearity is substituted with the swish activation~\cite{ramachandran2017searching} in some blocks.
The paper specifies the MobileNetV3small architecture for environments where resources are limited and the MobileNetV3large architecture for high-resource use cases. 
These architectures are very similar, with MobileNetV3large having more bottleneck blocks. 

\subsection{Side-channel analysis}
Side-channel analysis (SCA) exploits the physical leakages of electronic devices to extract secret information~\cite{kocher1996timing, kocher1999differential}. Such leakages could be power consumption, electromagnetic (EM) emanations, timing, optical, or sound, while secret information could be anything that has to remain confidential. The attacks based on side-channel information were first introduced in the 90's, targeting constrained firmly cryptographic devices such as smartcards~\cite{kocher1996timing, kocher1999differential} and they pose ever since a constant threat to the security of various embedded systems. In this work, we exploit the timing and EM side channels.

\subsubsection{Timing analysis}
Timing vulnerabilities in implementations arise from different sources, such as branching, cache hits/misses, and instructions. 
These vulnerabilities also pose a threat to cryptographic algorithms~\cite{kocher1996timing, page2002theoretical, bernstein2005cache}.
Timing attacks are typically based on the vulnerability of implementations where an operation takes a varying amount of time to complete, where this variation is due to the private key or other data being manipulated or even different instructions executed. 

\subsubsection{Power analysis}
Kocher et al. (1999)~\cite{kocher1999differential} introduced power consumption-based attacks called Simple Power Analysis (SPA) and Differential Power Analysis (DPA) by measuring the power consumption of microcontrollers during the execution of cryptographic algorithms. These attacks exploit the dependence between the power consumption of a device and the executed operations and processed data by the device. 
SPA is a method to visually analyze and interpret the collected power consumption measurements, also called \textit{traces}. It often requires a few or just a single trace to extract information about the operations and data used in the targeted algorithm. DPA exploits the dependency of power consumption on the processed data. The small variations in power due to different data being processed can allow an adversary to extract secret information (e.g., secret key) about the targeted algorithm using power measurements. 

\subsubsection{Electromagnetic emanations}
Electromagnetic (EM) emanations have been exploited since the Second World War~\cite{singh1999code} and pose a massive security issue for sensitive systems. Wim van Eck~\cite{van1985electromagnetic} was the first to publish a paper about the risk of information leakage due to electromagnetic radiation using equipment that anyone can acquire. His work demonstrated the danger of EM radiation by reconstructing the frames of the video from display units using EM radiation.

Since then, EM radiation has also been used to break cryptographic implementations~\cite{QS01, kuhn1998soft} or eavesdrop on display units~\cite{elibol2012realistic, hongxin2009recognition, LiuSW0LBL21}. Similar to power analysis, Simple EM Analysis (SEMA) and Differential EM Analysis (DEMA) are methods that work exactly the same way as their counterparts in power analysis, with the exception of the traces consisting of EM measurements. In this work, we use electromagnetic emanations in combination with timing information to distinguish the architectures.

\section{Architecture extraction} \label{archext}
\subsection{Threat model}

\begin{table}[t!]
\begin{center}
\begin{tabular}{ |c|c| } 
 \hline
 Name& \# parameters \\
 \hline
 \hline
 EfficientNetB0 & 5.3 M \\ 
 \hline
 EfficientNetB1 & 7.9 M \\ 
 \hline
 EfficientNetB2 & 9.2 M \\ 
 \hline 
 EfficientNetB3 & 12.3 M \\ 
 \hline
 EfficientNetB4 & 19.5 M \\ 
 \hline
 EfficientNetB5 & 30.6 M \\ 
 \hline 
 EfficientNetB6 & 43.3 M \\ 
 \hline
 MobileNet & 4.3 M \\ 
 \hline
 MobileNetv2 & 3.5 M \\ 
 \hline
 MobileNetv3small & 2.5 M \\ 
 \hline
 MobileNetv3large & 5.4 M \\ 
 \hline
 Densenet121 & 8.1 M \\ 
 \hline
 Densenet169 & 14.3 M \\ 
 \hline
 Densenet201 & 20.2 M \\ 
 \hline
 NASNetMobile & 5.3 M \\ 
 \hline

\end{tabular}
\end{center}
\caption[]{Analyzed convolutional neural network architectures}
\label{table:1}
\end{table}

In our threat model, the adversary has the following knowledge and capabilities.

\begin{itemize}
    \item[\textbf{A1:}] Physical access to the target device.
    \item[\textbf{A2:}] Access to an identical device for profiling.
    \item[\textbf{A3:}] Capability to collect electromagnetic side-channel measurements. 
    \item[\textbf{A4:}] Knowledge that one of the 15 architectures listed in Table \ref{table:1} is executed on the target device. 
\end{itemize}

The capabilities \textbf{A1}-\textbf{A3} are standard assumptions in profiled side-channel attacks~\cite{picek2022sok}. The assumption of \textbf{A1} can be relaxed as the adversary requires limited amount of physical access to the target device because the attack requires only a single trace to identify the correct architecture. In addition, the assumption of \textbf{A4} is motivated by the investigated architectures' efficiency in resource-constrained environments like embedded devices. Furthermore, developers may choose to pick an off-the-shelf architecture that is proven to work instead of developing custom architectures, which can be an expensive and time-consuming process. In our experiments, we use the same device for profiling and attacking.

\subsection{NVIDIA neural network implementations}
In our attack, we are considering the implementations from NVIDIA's TensorRT deep learning inference framework. 
TensorRT is a library written by NVIDIA to support deep learning inference by running neural networks efficiently and quickly on NVIDIA hardware. 

TensorRT works as follows: 
\begin{enumerate}
    \item the user defines the neural network model
    \item the user defines the desired optimizations for the model
    \item TensorRT builds an engine based on the defined model and desired optimizations
\end{enumerate}

Optimizations include layer fusions and calibration of the precision of calculations. Given the precision constraints, TensorRT times different implementations and chooses the fastest ones for the model. The built engine includes layer implementations and model weights, which can be subsequently used for inference. In our experiments, we restricted the models to use implementations with half-precision calculations to decrease the memory footprint of the architectures as some of them require more than the available DRAM in the device if single-precision calculations are used.
\begin{figure}[!t]
    \centering
    \includegraphics[width=.8\textwidth]{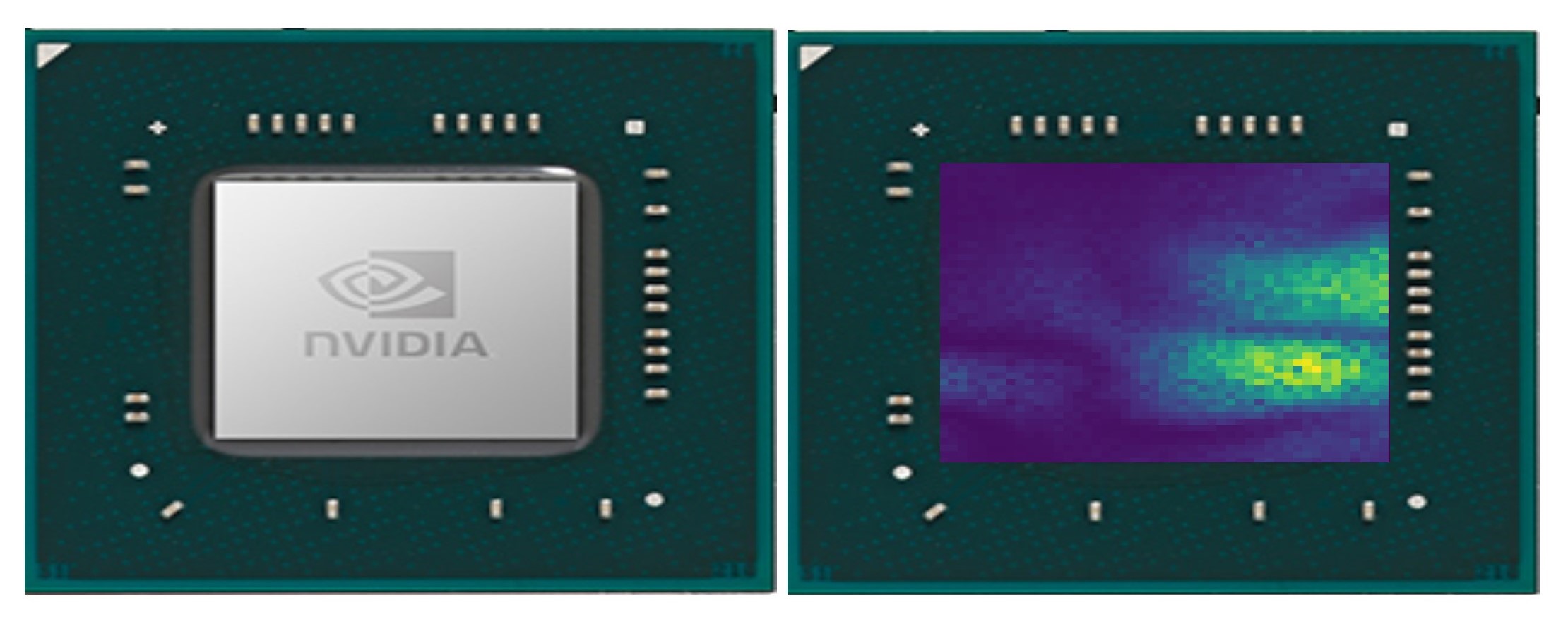}
    \caption{Heatmap of 78MHz clock frequency after scanning the chip of the Jetson Nano device. The heatmap was generated by applying the Fourier-transform on traces collected at each point on the chip. Purple indicates no activity of the 78MHz clock frequency while yellow indicates the highest activity of this frequency at a certain point. Multiple yellow points can be used to mount a successful architecture extraction attack.}
    \label{fig::heatmap}
\end{figure}

\subsection{Measurement collection}
\begin{figure}[!t]
    \centering
    \includegraphics[width=.8\textwidth]{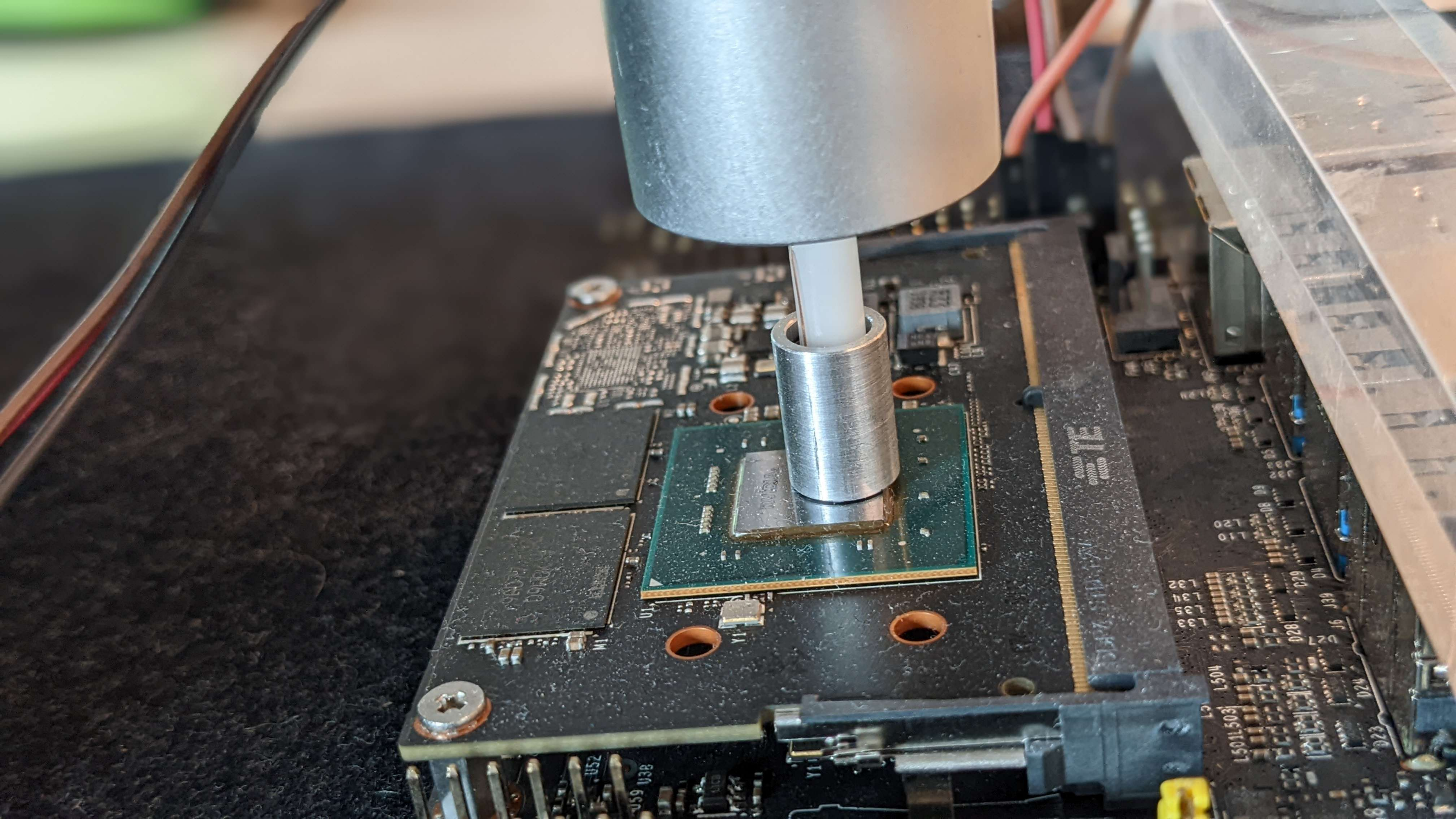}
    \caption{Location of the Riscure EM probe. The probe tip is located above the chip.}
    \label{fig::Probe and device}
\end{figure}

We use the PicoScope 3207B oscilloscope with a Riscure EM probe~\cite{riscure_probe} and Riscure EM probe station~\cite{riscure_probe} to collect electromagnetic side-channel measurements in the architecture extraction attack. 
In order to capture the inference of the neural networks from the device, a GPIO pin on the Jetson Nano's board is used as a trigger for the oscilloscope to signal when the inference operation is about to start.
In the architecture extraction experiment, we set the GPU cores of the device to operate at 76 MHz clock frequency, so we set the sampling rate of the oscilloscope at 1GS/s.
In order to detect where the chip of the Jetson Nano leaks the most information, the whole chip was scanned. The results of the scan are shown in Figure \ref{fig::heatmap}. Based on the figure and experiments, there are multiple locations where the architectural information of neural networks leaks.
The final location of the probe is shown in Figure \ref{fig::Probe and device}.

\begin{figure}[!t]
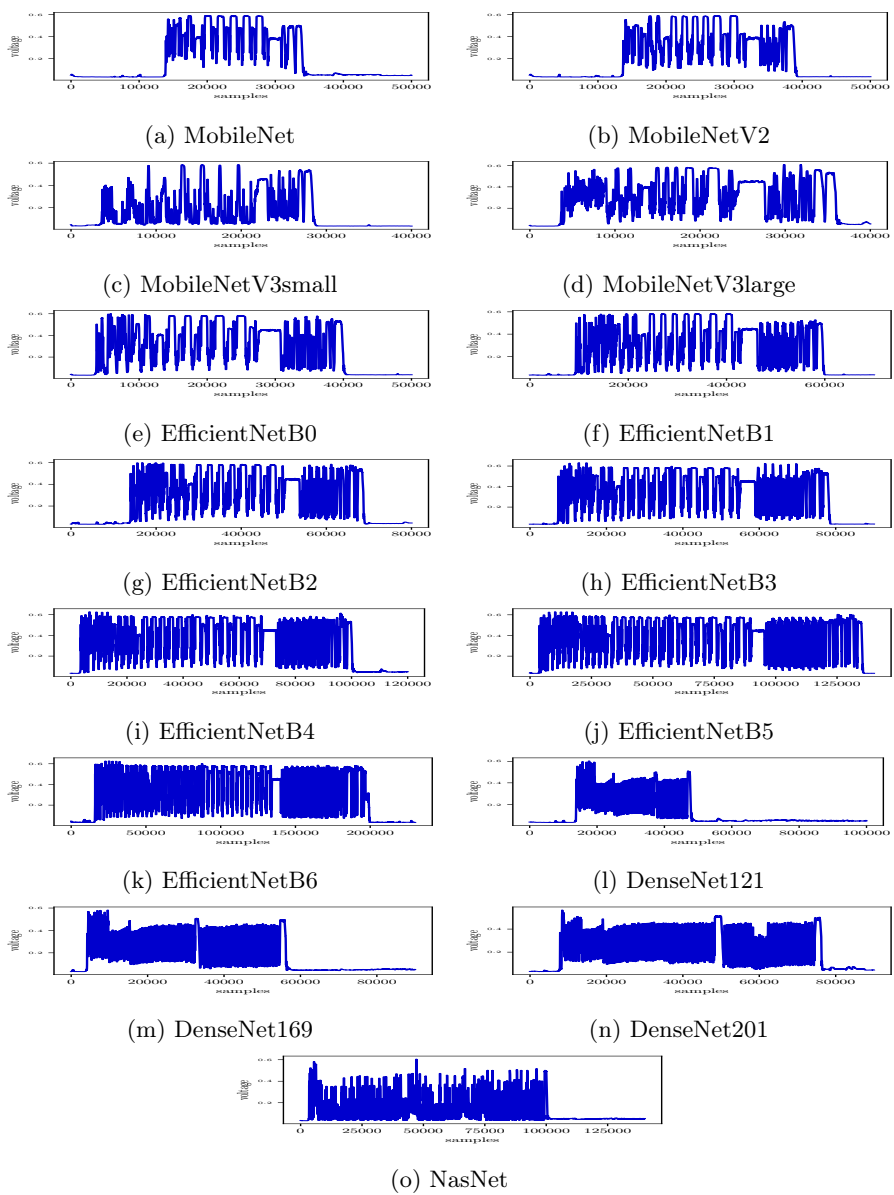

    \centering
    \subfloat[][MobileNet]{%
    \resizebox{.5\textwidth}{.07\textheight}{\input{img/networks/nn2_1.pgf}}}
    \subfloat[][MobileNetV2]{%
    \resizebox{.5\textwidth}{.07\textheight}{\input{img/networks/nn2_0.pgf}}}

    \subfloat[][MobileNetV3small]{%
    \resizebox{.5\textwidth}{.07\textheight}{\input{img/networks/nn2_2.pgf}}}
    \subfloat[][MobileNetV3large]{%
    \resizebox{.5\textwidth}{.07\textheight}{\input{img/networks/nn2_3.pgf}}}

    \subfloat[EfficientNetB0]{%
    \resizebox{.5\textwidth}{.07\textheight}{\input{img/networks/nn2_4.pgf}}}
    \subfloat[EfficientNetB1]{%
    \resizebox{.5\textwidth}{.07\textheight}{\input{img/networks/nn2_5.pgf}}}

    \subfloat[EfficientNetB2]{%
    \resizebox{.5\textwidth}{.07\textheight}{\input{img/networks/nn2_6.pgf}}}
    \subfloat[EfficientNetB3]{%
    \resizebox{.5\textwidth}{.07\textheight}{\input{img/networks/nn2_7.pgf}}}

    \subfloat[EfficientNetB4]{%
    \resizebox{.5\textwidth}{.07\textheight}{\input{img/networks/nn2_8.pgf}}}
    \subfloat[EfficientNetB5]{%
    \resizebox{.5\textwidth}{.07\textheight}{\input{img/networks/nn2_9.pgf}}}

    \subfloat[EfficientNetB6]{%
    \resizebox{.5\textwidth}{.07\textheight}{\input{img/networks/nn2_10.pgf}}}
    \subfloat[DenseNet121]{%
    \resizebox{.5\textwidth}{.07\textheight}{\input{img/networks/nn2_11.pgf}}}

    \subfloat[DenseNet169]{%
    \resizebox{.5\textwidth}{.07\textheight}{\input{img/networks/nn2_12.pgf}}}
    \subfloat[DenseNet201]{%
    \resizebox{.5\textwidth}{.07\textheight}{\input{img/networks/nn2_13.pgf}}}

    \subfloat[NasNet]{%
    \resizebox{.5\textwidth}{.07\textheight}{\input{img/networks/nn2_14.pgf}}}
    \caption{Example traces of the investigated architectures.}\label{fig::all_arch}

\end{figure}
\newpage
\subsection{Architecture extraction using SEMA and timing analysis}
Here we discuss Simple EM Analysis and timing analysis on the collected traces using. The traces shown in this section are not the raw traces but the preprocessed versions of those. 
For preprocessing, we applied windowed averaging of size 1\,000 on the absolute value of the measurements. Alignment of the traces is not required for this attack.

Figure \ref{fig::all_arch} shows example traces of the architectures that are investigated in this paper.
For the MobileNet and MobileNetv2 architectures, there are clear timing differences between them, showing the MobileNetV2 takes more time to execute. 
According to the benchmarks in the original paper~\cite{sandler2018mobilenetv2}, MobileNetV2 is faster than MobileNetV1. However, the experiments in the original paper were carried out on the CPU of the Google Pixel 1 smartphone, using TensorFlow Lite, so this might explain the difference.
The displayed patterns are similar for the architectures, which is expected as MobileNetV2's building block is based on MobileNet's building block. 
Regarding the MobileNetV3small and MobileNetV3large architectures, the execution time for the MobilNetV3large architecture is substantially larger than that of MobileNetV3small, as expected.
However, the execution time for the MobileNetV3small architecture is very similar to that of the MobileNetV2.
The DenseNet 121, DenseNet169, and DenseNet201 architectures display very different EM patterns than the rest of the architectures. In addition, the displayed patterns are very similar when compared to each other.
However, the timing differences clearly identify the correct architecture DenseNet architecture.  
Regarding the EfficientNet architectures, the EM patterns are similar to that of the MobileNet architectures, as expected, but the timing differences give away the correct architecture.
Lastly, the EM patterns shown by the NasNet architecture are quite distinct compared to the previous architectures. In terms of execution times, NasNet is very similar to EfficientNetB4, but NasNet's EM amplitude frequently drops near zero. 

\subsection{Architecture extraction using deep learning}

In this section, we present how the architecture extraction attack can be automated using deep learning by framing the problem as a classification problem.
The models for each architecture for training were created using TensorFlow.
For some architectures, the TensorFlow implementation involves preprocessing layers that actually do not belong to the architecture. These preprocessing layers make it possible for the network to receive inputs that are not preprocessed. These layers were removed before creating the models so that every architecture uniformly does not have preprocessing layers. Besides this, the default parameter values of the TensorFlow implementations were used. 

To train and validate the deep learning classifier, $n=5$ models $M_{a, i}$ were created for every architecture ($g=15$). For testing, $t=3$ models were created per architecture. These only differ in their weights $W_{a, i}$ as all the weights are sampled randomly from a normal distribution (with mean 0 and variance 1) for every model. 

Formally, 
$$M_{a, i} = f_{a}(x; W_{a, i})     (i=1,\ldots, n+t; a=1, \ldots,g)$$
which means that $i$'th model for the $a$-th architecture is defined as a function of its weights and its inputs ($x$). Additionally, 
$$W_{a, i} \neq W_{a, j}$$
where $ j \in \{1, 2, .., n+t\} \setminus \{i\}$. The function $f_{a}$ depends on the architecture of the model.  
Overall, $g \times (n+t) = 120$ models were created. The input and batch size of the models during the experiments were set to 32x32x3 and 1, respectively.

\input{classifier.tex}
We define a simple convolutional neural network as our classifier, shown in Table \ref{layers_table}, as they have proven to be effective in the SCA context~\cite{kim2019make, picek2022sok}.
We collect 200 measurements for each model in the training and validation sets and 20 measurements per model in the test set, which amounts to 15\,900 traces altogether.
Out of the 15\,900 traces, the test set contains 900 measurements, and the remaining 15\,000 traces are divided into training and validation sets in a 70:30 ratio, i.e., the model is trained using 10\,500 traces and validated using 4\,500 traces. 
In addition, early stopping is used to avoid overfitting. After the model is trained, it is evaluated on the test set and the accuracy of the model was 99\%. 
Note from the previous section that distinguishing the architectures is not difficult; hence, the almost perfect accuracy is not surprising. 

\section{Discussion}\label{discussion}
\subsection{Limitations}
The attacks described are specific to this device's GPU and the CUDA kernel implementations provided by the specific TensorRT version used. In addition, we work with the assumption that well-known architectures are used by the victim. If a target device runs a different architecture, that is not in the dataset used for profiling, then the attack does not work unless profiling is extended to more architectures. However, with extensive profiling, we believe it is possible to cover a wide array of architectures with different types of layers.

\subsection{Mitigation}
Traditional ways to contain electromagnetic radiation, such as proper shielding or introducing noise to decrease the Signal-to-Noise ratio, could alleviate the problem~\cite{mangard2008power}. 

Additionally, the architectures investigated in this work are popular because of their efficiency and accuracy. However, ignoring these architectures and designing custom networks could make an adversary's job significantly harder. A custom-designed neural network basically means an infinite number of possible combinations of layers, layer sizes, etc. On the other hand, there are common design principles for neural networks which narrow down the search space. For instance, if a neural network performs classification, then it is safe to assume that the last layer has a softmax activation. 

Profiling also applies to custom-made neural networks, and a persistent adversary could make a comprehensive profile that could also identify the types of layers and layer sizes, as these are the main factors that influence EM measurements.

\subsection{Alternative method}
In this work, analyzing traces of whole architectures is enough to show that reverse engineering the architecture is possible.
 However, one could reverse engineer a whole architecture by running just parts of the architecture on the unprotected device. In other words, starting with only the first layer of the architecture, then with the first two, then the first three, and so on.  With this perhaps a bit of a time-consuming (due to the large number of layers in the investigated architectures) method, the individual layers can be identified in the traces, not just the whole architecture. Since the number of parameters for these architectures is constant, except perhaps for the first and last layers, this method remains viable. The traces for the first and last layers can potentially be different because input and output sizes are specific to each problem.

\subsection{Example of breaking down network}
In order to identify individual layers, one has to consider the layer type as well as the activation (if any) of the layer. As we have seen in the classification results, different weights barely impact the overall EM trace. Thus, we can concentrate on building templates for one-layer MLPs with and without activation, 2-layer MLPs with and without activation, and so on. 
To that end, a 3-layer MLP will be reverse-engineered using this method. In the experiment, the input batch size is 1, the input size is 100, and every fully connected layer has 32 neurons.

\begin{figure}[t!]
    \begin{center}
        \scalebox{.7}{\input{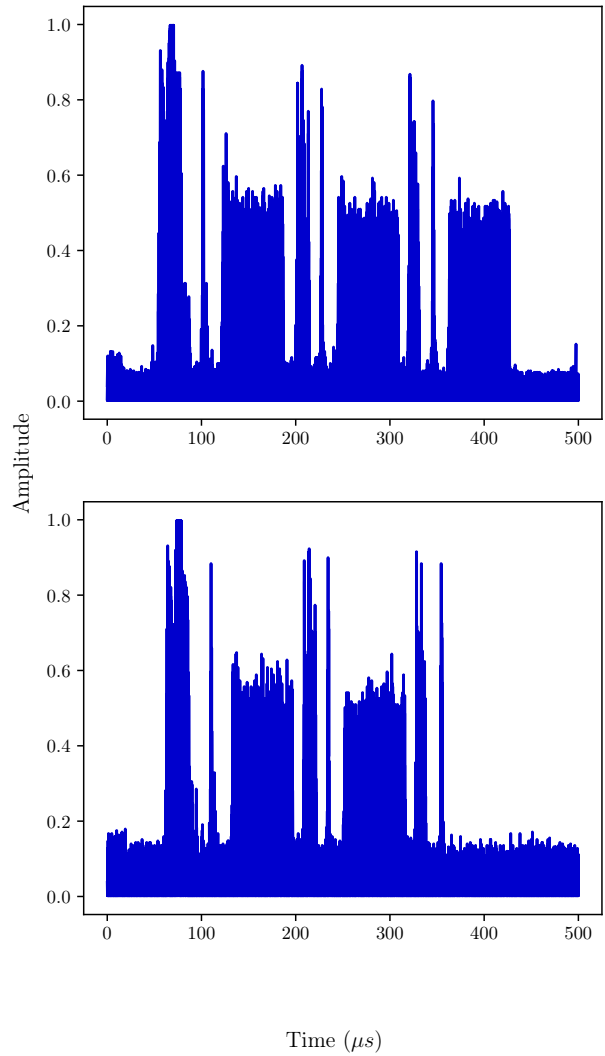}}
    \end{center}
    \caption{3-layer MLP with 3 ReLU activations (top) and 2 ReLU activations (bottom)}
    \label{fig::3mlp_2ReLU_3ReLU}
\end{figure}

\begin{figure}[t!]
    \begin{center}
        \scalebox{.7}{\input{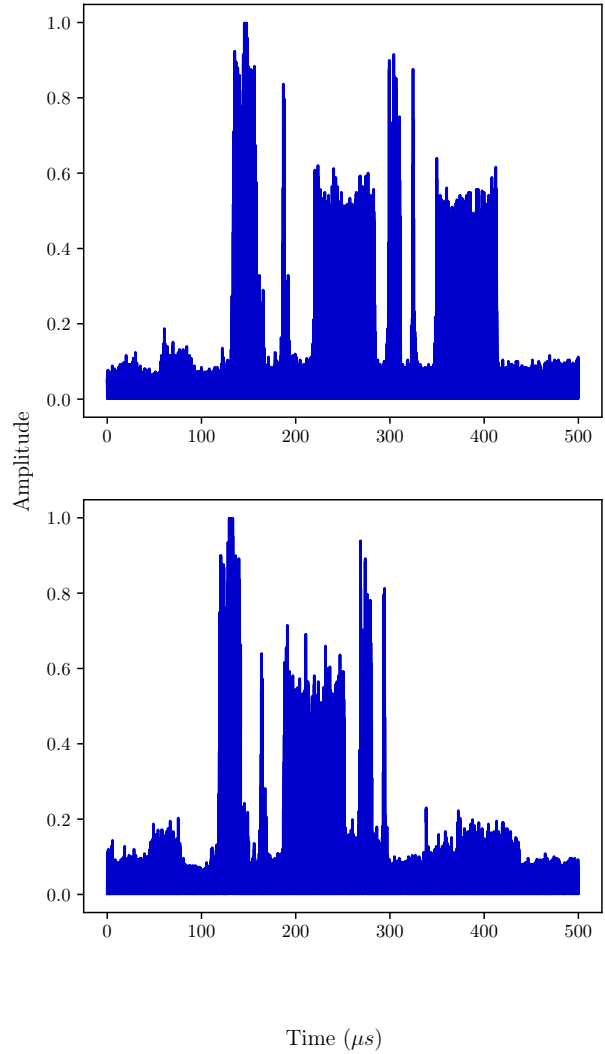}}
    \end{center}
    \caption{2-layer MLP with 2 ReLU activations (top) and 1 ReLU activation (bottom)}
    \label{fig::2mlp_1ReLU_2ReLU}
\end{figure}

\begin{figure}[t!]
    \begin{center}
        \scalebox{.7}{\input{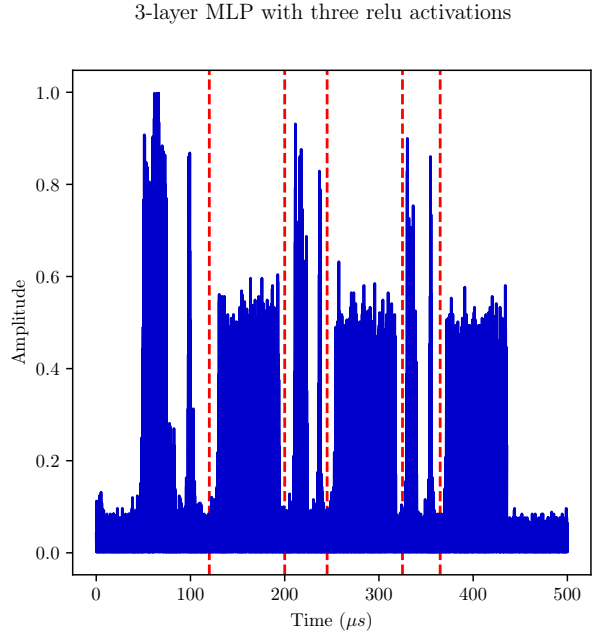}}
    \end{center}
    \caption{3-layer MLP with boundaries between fully connected layers and activations.}
    \label{fig::3mlp}
\end{figure}

Figure \ref{fig::3mlp_2ReLU_3ReLU} shows how the EM trace changes if a ReLU activation is removed. The top figure is a trace of a 3-layer MLP where the fully connected layers are followed by ReLU activation. The bottom figure is a trace of the same MLP, except that a ReLU layer does not follow the last fully connected layer.

Continuing the removal of layers and activations one by one, the top figure in Figure \ref{fig::2mlp_1ReLU_2ReLU} shows how the EM trace changes if a fully connected layer is removed. The MLP in the bottom figure in Figure \ref{fig::3mlp_2ReLU_3ReLU} has a third fully connected layer, and the top figure in Figure \ref{fig::2mlp_1ReLU_2ReLU} is the same  MLP except that the last fully connected layer is missing.
Next, the removal of the ReLU activation that follows the second fully connected layer leads to a trace as that of the bottom figure in Figure \ref{fig::2mlp_1ReLU_2ReLU}.

 Removing layers one by one helps identify layer boundaries. Figure \ref{fig::3mlp} shows the trace for the 3-layer MLP with boundaries drawn with red dashed lines after every fully connected layer and every activation. Overall, profiling can also be executed on a more granular level, e.g. layer level, but this requires more profiling to cover all the possible layer types with varying sizes.
 
\section{Conclusions}\label{conclusions}
In this paper, the susceptibility of neural networks to side-channel attacks was analyzed on NVIDIA Jetson Nano. The neural networks ran on the GPU of the device, which is a commonly chosen platform for real-world neural network implementations. 

In our attack, popular convolutional neural network architectures were classified based on the EM side channel. The chosen architectures are a common choice in practice, especially in embedded devices, when the size and latency of the network are important as resources are limited. The results show that the analyzed architectures are easily distinguishable from each other, and this process can be automated using a deep learning classifier. 

Overall, the neural network implementations of NVIDIA's TensorRT framework are vulnerable to architecture extraction using side-channel attacks despite the networks running in a highly parallel and noisy environment.

%
%
%
\bibliographystyle{splncs04}
\bibliography{sample}
%

\end{document}

%% file: classifier.tex
\begin{table}[t!]
\begin{center}
\resizebox{\linewidth}{!}{
\begin{tabular}{ llc } 
 \toprule
 \textbf{Layer type}& \textbf{Hyperparameters} &  \textbf{Activation} \\
 \midrule
 Conv1D & filters: 32, kernel-size: 500, strides: 50 &  ReLU\\ 
 Conv1D & filters: 32, kernel-size: 300, strides: 10 &  ReLU\\ 
 Max-pool & pool-size: 10, strides: 5  &  -\\ 
 Flatten & -  &  - \\ 
 Dense & neurons: 32  &  ReLU \\ 
 Dropout & dropout rate: 0.2 & - \\ 
 Dense & neurons: 15  &  softmax\\ 
 \bottomrule
\end{tabular}
}
\end{center}
\caption{Classifier architecture and hyperparameters}
\label{layers_table}
\end{table}

%% file: main.bbl
\begin{thebibliography}{10}
\providecommand{\url}[1]{\texttt{#1}}
\providecommand{\urlprefix}{URL }
\providecommand{\doi}[1]{https://doi.org/#1}

\bibitem{riscure_probe}
\url{https://web.archive.org/web/20220119062522/https://www.riscure.com/uploads/2017/07/inspector_brochure.pdf},
  accessed: 2022-01-25

\bibitem{googletranslate}
Google translate research.
  \url{https://ai.googleblog.com/2020/06/recent-advances-in-google-translate.html}

\bibitem{batina2019csi}
Batina, L., Bhasin, S., Jap, D., Picek, S.: {CSI}--{NN}: Reverse engineering of
  neural network architectures through electromagnetic side channel. In: 28th
  {USENIX} Security Symposium {USENIX} Security 19). pp. 515--532 (2019)

\bibitem{bernstein2005cache}
Bernstein, D.J.: Cache-timing attacks on {AES}  (2005)

\bibitem{chmielewski2021reverse}
Chmielewski, {\L}., Weissbart, L.: On reverse engineering neural network
  implementation on {GPU}. In: International Conference on Applied Cryptography
  and Network Security. pp. 96--113. Springer (2021)

\bibitem{chollet2017xception}
Chollet, F.: Xception: {D}eep learning with depthwise separable convolutions.
  In: Proceedings of the IEEE conference on computer vision and pattern
  recognition. pp. 1251--1258 (2017)

\bibitem{elibol2012realistic}
Elibol, F., Sarac, U., Erer, I.: Realistic eavesdropping attacks on computer
  displays with low-cost and mobile receiver system. In: 2012 Proceedings of
  the 20th European Signal Processing Conference (EUSIPCO). pp. 1767--1771.
  IEEE (2012)

\bibitem{he2016deep}
He, K., Zhang, X., Ren, S., Sun, J.: Deep residual learning for image
  recognition. In: Proceedings of the IEEE conference on computer vision and
  pattern recognition. pp. 770--778 (2016)

\bibitem{hongxin2009recognition}
Hongxin, Z., Yuewang, H., Jianxin, W., Yinghua, L., Jinling, Z.: Recognition of
  electro-magnetic leakage information from computer radiation with {SVM}.
  Computers \& Security  \textbf{28}(1-2),  72--76 (2009)

\bibitem{howard2019searching}
Howard, A., Sandler, M., Chu, G., Chen, L.C., Chen, B., Tan, M., Wang, W., Zhu,
  Y., Pang, R., Vasudevan, V., et~al.: Searching for {M}obilenet{v}3. In:
  Proceedings of the IEEE/CVF International Conference on Computer Vision. pp.
  1314--1324 (2019)

\bibitem{howard2017mobilenets}
Howard, A.G., Zhu, M., Chen, B., Kalenichenko, D., Wang, W., Weyand, T.,
  Andreetto, M., Adam, H.: Mobilenets: {E}fficient convolutional neural
  networks for mobile vision applications. arXiv preprint arXiv:1704.04861
  (2017)

\bibitem{hu2018squeeze}
Hu, J., Shen, L., Sun, G.: Squeeze-and-excitation networks. In: Proceedings of
  the IEEE conference on computer vision and pattern recognition. pp.
  7132--7141 (2018)

\bibitem{huang2017densely}
Huang, G., Liu, Z., Van Der~Maaten, L., Weinberger, K.Q.: Densely connected
  convolutional networks. In: Proceedings of the IEEE conference on computer
  vision and pattern recognition. pp. 4700--4708 (2017)

\bibitem{kim2019make}
Kim, J., Picek, S., Heuser, A., Bhasin, S., Hanjalic, A.: Make some noise.
  unleashing the power of convolutional neural networks for profiled
  side-channel analysis. IACR Transactions on Cryptographic Hardware and
  Embedded Systems pp. 148--179 (2019)

\bibitem{kocher1999differential}
Kocher, P., Jaffe, J., Jun, B.: Differential power analysis. In: Annual
  international cryptology conference. pp. 388--397. Springer (1999)

\bibitem{kocher1996timing}
Kocher, P.C.: Timing attacks on implementations of {D}iffie-{H}ellman, {RSA},
  {DSS}, and other systems. In: Annual International Cryptology Conference. pp.
  104--113. Springer (1996)

\bibitem{krizhevsky2012imagenet}
Krizhevsky, A., Sutskever, I., Hinton, G.E.: Imagenet classification with deep
  convolutional neural networks. Advances in neural information processing
  systems  \textbf{25},  1097--1105 (2012)

\bibitem{kuhn1998soft}
Kuhn, M.G., Anderson, R.J.: Soft tempest: Hidden data transmission using
  electromagnetic emanations. In: International Workshop on Information Hiding.
  pp. 124--142. Springer (1998)

\bibitem{clairvoyance}
Liang, S., Zhan, Z., Yao, F., Cheng, L., Zhang, Z.: Clairvoyance: {E}xploiting
  {Far}-field {EM} {E}manations of {GPU} to "{S}ee" {Y}our {DNN} {M}odels
  through {O}bstacles at a {D}istance. In: 2022 IEEE Security and Privacy
  Workshops (SPW). pp. 312--322 (2022). \doi{10.1109/SPW54247.2022.9833894}

\bibitem{lin2013network}
Lin, M., Chen, Q., Yan, S.: Network in network. arXiv preprint arXiv:1312.4400
  (2013)

\bibitem{liu2020deep}
Liu, L., Ouyang, W., Wang, X., Fieguth, P., Chen, J., Liu, X., Pietik{\"a}inen,
  M.: Deep learning for generic object detection: {A} survey. International
  journal of computer vision  \textbf{128}(2),  261--318 (2020)

\bibitem{LiuSW0LBL21}
Liu, Z., Samwel, N., Weissbart, L., Zhao, Z., Lauret, D., Batina, L., Larson,
  M.A.: Screen {G}leaning: {A} {S}creen {R}eading {TEMPEST} {A}ttack on
  {M}obile devices {E}xploiting an {E}lectromagnetic {S}ide {C}hannel. In: 28th
  Annual Network and Distributed System Security Symposium, {NDSS} 2021,
  virtually, February 21-25, 2021. The Internet Society (2021),
  \url{https://www.ndss-symposium.org/ndss-paper/screen-gleaning-a-screen-reading-tempest-attack-on-mobile-devices-exploiting-an-electromagnetic-side-channel/}

\bibitem{DBLP:conf/uss/MaiaXLGZ22}
Maia, H.T., Xiao, C., Li, D., Grinspun, E., Zheng, C.: Can one hear the shape
  of a neural network?: Snooping the {GPU} via magnetic side channel. In:
  Butler, K.R.B., Thomas, K. (eds.) 31st {USENIX} Security Symposium, {USENIX}
  Security 2022, Boston, MA, USA, August 10-12, 2022. pp. 4383--4400. {USENIX}
  Association (2022),
  \url{https://www.usenix.org/conference/usenixsecurity22/presentation/maia}

\bibitem{mangard2008power}
Mangard, S., Oswald, E., Popp, T.: Power analysis attacks: Revealing the
  secrets of smart cards, vol.~31. Springer Science \& Business Media (2008)

\bibitem{openai2023gpt4}
OpenAI: Gpt-4 technical report (2023). \doi{10.48550/arXiv.2303.08774}

\bibitem{otter2020survey}
Otter, D.W., Medina, J.R., Kalita, J.K.: A survey of the usages of deep
  learning for natural language processing. IEEE transactions on neural
  networks and learning systems  \textbf{32}(2),  604--624 (2020)

\bibitem{page2002theoretical}
Page, D.: Theoretical use of cache memory as a cryptanalytic side-channel.
  Cryptology ePrint Archive  (2002)

\bibitem{picek2022sok}
Picek, S., Perin, G., Mariot, L., Wu, L., Batina, L.: So{K}: {D}eep
  learning-based physical side-channel analysis. ACM Computing Surveys  (2022)

\bibitem{QS01}
Quisquater, J.J., Samyde, D.: {{ElectroMagnetic Analysis (EMA): Measures and
  Counter-Measures for Smard Cards}}. In: Attali, I., Jensen, T.P. (eds.) Smart
  Card Programming and Security (E-smart 2001). Lecture Notes in Computer
  Science, vol.~2140, pp. 200--210. Springer-Verlag (2001)

\bibitem{ramachandran2017searching}
Ramachandran, P., Zoph, B., Le, Q.V.: Searching for activation functions. arXiv
  preprint arXiv:1710.05941  (2017)

\bibitem{sandler2018mobilenetv2}
Sandler, M., Howard, A., Zhu, M., Zhmoginov, A., Chen, L.C.: Mobilenetv2:
  {I}nverted residuals and linear bottlenecks. In: Proceedings of the IEEE
  conference on computer vision and pattern recognition. pp. 4510--4520 (2018)

\bibitem{silver2016mastering}
Silver, D., Huang, A., Maddison, C.J., Guez, A., Sifre, L., Van Den~Driessche,
  G., Schrittwieser, J., Antonoglou, I., Panneershelvam, V., Lanctot, M.,
  et~al.: Mastering the game of go with deep neural networks and tree search.
  nature  \textbf{529}(7587),  484--489 (2016)

\bibitem{silver2017mastering}
Silver, D., Hubert, T., Schrittwieser, J., Antonoglou, I., Lai, M., Guez, A.,
  Lanctot, M., Sifre, L., Kumaran, D., Graepel, T., et~al.: Mastering chess and
  shogi by self-play with a general reinforcement learning algorithm. arXiv
  preprint arXiv:1712.01815  (2017)

\bibitem{simonyan2014very}
Simonyan, K., Zisserman, A.: Very deep convolutional networks for large-scale
  image recognition. arXiv preprint arXiv:1409.1556  (2014)

\bibitem{singh1999code}
Singh, S.: The code book, vol.~7. Doubleday New York (1999)

\bibitem{tan2019efficientnet}
Tan, M., Le, Q.: Efficientnet: Rethinking model scaling for convolutional
  neural networks. In: International conference on machine learning. pp.
  6105--6114. PMLR (2019)

\bibitem{van1985electromagnetic}
Van~Eck, W.: Electromagnetic radiation from video display units: An
  eavesdropping risk? Computers \& Security  \textbf{4}(4),  269--286 (1985)

\bibitem{vaswani2017attention}
Vaswani, A., Shazeer, N., Parmar, N., Uszkoreit, J., Jones, L., Gomez, A.N.,
  Kaiser, {\L}., Polosukhin, I.: Attention is all you need. Advances in neural
  information processing systems  \textbf{30} (2017)

\bibitem{vinyals2019grandmaster}
Vinyals, O., Babuschkin, I., Czarnecki, W.M., Mathieu, M., Dudzik, A., Chung,
  J., Choi, D.H., Powell, R., Ewalds, T., Georgiev, P., et~al.: Grandmaster
  level in {S}tar{C}raft {II} using multi-agent reinforcement learning. Nature
  \textbf{575}(7782),  350--354 (2019)

\bibitem{9000972}
Xiang, Y., Chen, Z., Chen, Z., Fang, Z., Hao, H., Chen, J., Liu, Y., Wu, Z.,
  Xuan, Q., Yang, X.: Open {DNN} {B}ox by {P}ower {S}ide-{C}hannel {A}ttack.
  IEEE Transactions on Circuits and Systems II: Express Briefs
  \textbf{67}(11),  2717--2721 (2020). \doi{10.1109/TCSII.2020.2973007}

\bibitem{xie2017aggregated}
Xie, S., Girshick, R., Doll{\'a}r, P., Tu, Z., He, K.: Aggregated residual
  transformations for deep neural networks. In: Proceedings of the IEEE
  conference on computer vision and pattern recognition. pp. 1492--1500 (2017)

\bibitem{zhang2018shufflenet}
Zhang, X., Zhou, X., Lin, M., Sun, J.: Shufflenet: An extremely efficient
  convolutional neural network for mobile devices. In: Proceedings of the IEEE
  conference on computer vision and pattern recognition. pp. 6848--6856 (2018)

\bibitem{zoph2018learning}
Zoph, B., Vasudevan, V., Shlens, J., Le, Q.V.: Learning transferable
  architectures for scalable image recognition. In: Proceedings of the IEEE
  conference on computer vision and pattern recognition. pp. 8697--8710 (2018)

\end{thebibliography}
